%% file: ms.tex
\pdfoutput=1
\documentclass{article}

\usepackage[utf8]{inputenc}
\usepackage{graphicx}
\usepackage{pdfpages}
\usepackage{color}   
\usepackage{hyperref}
\hypersetup{
    colorlinks=true, 
    linktoc=all,     
    linkcolor=blue,  
    urlcolor= blue,
}
\usepackage{biblatex} 
\usepackage[english]{babel}
\usepackage{csquotes}

\begin{document}

\includepdf[pages=-]{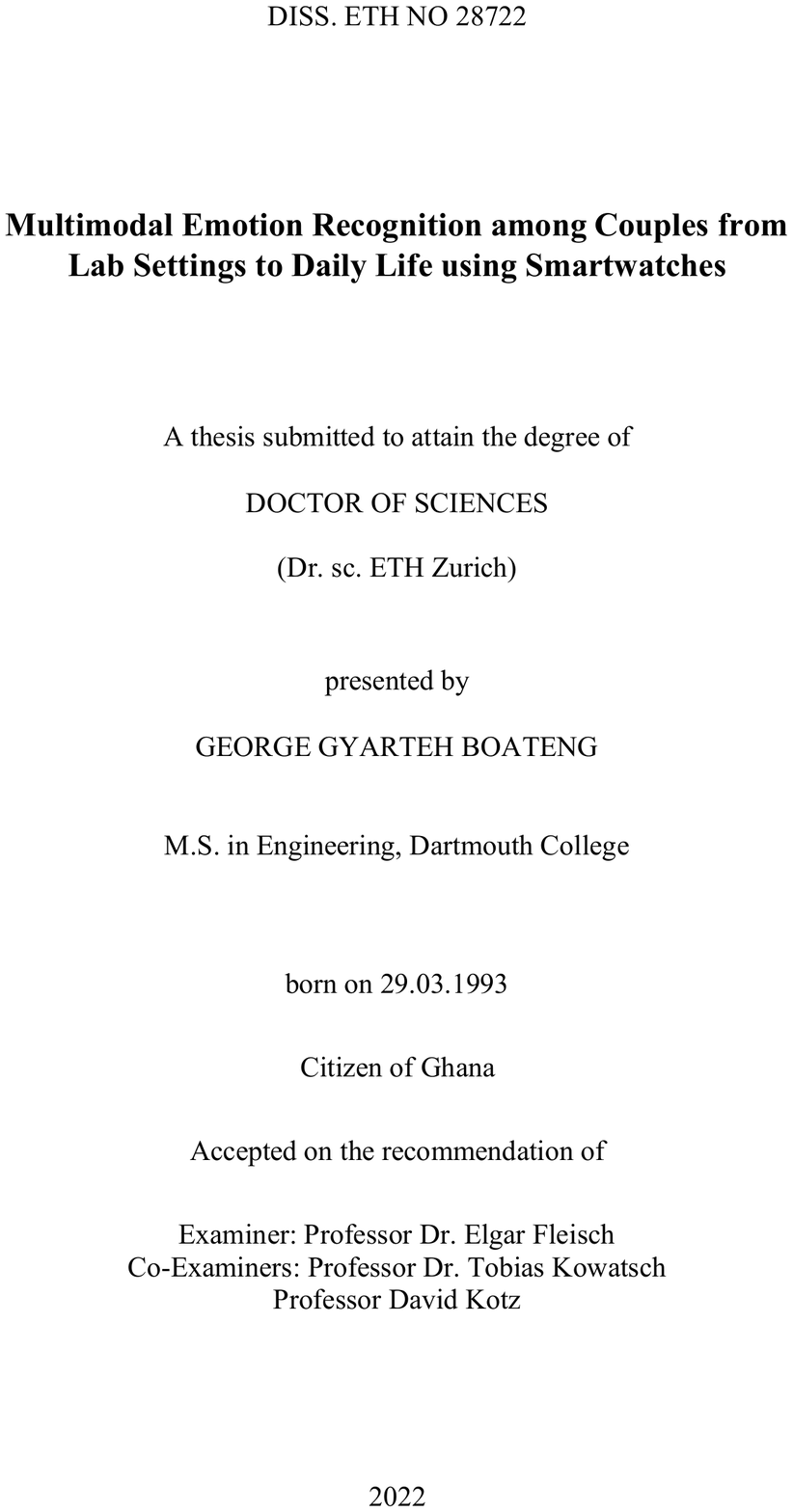}

\pagenumbering{roman}
\clearpage
\textit{“In God we trust. All others must bring (raw) data.”}  \bigbreak
\hfill W. Edwards Deming

\clearpage
\section*{Summary}
\addcontentsline{toc}{section}{Summary}
Couples generally manage chronic diseases together and the management takes an emotional toll on both patients and their romantic partners. Consequently, recognizing the emotions of each partner in daily life could provide an insight into their emotional well-being in chronic disease management. The emotions of partners are currently inferred in the lab and daily life using self-reports which are not practical for continuous emotion assessment or observer reports which are manual, time-intensive, and costly. Currently, there exists no comprehensive overview of works on emotion recognition among couples. Furthermore, approaches for emotion recognition among couples have (1) focused on English-speaking couples in the U.S., (2) used data collected from the lab, and (3) performed recognition using observer ratings rather than partner’s self-reported / subjective emotions. 

In this body of work contained in this thesis (8 papers — 5 published and 3 currently under review in various journals), we fill the current literature gap on couples’ emotion recognition, develop emotion recognition systems using 161 hours of data from a total of 1,051 individuals, and make contributions towards taking couples’ emotion recognition from the lab which is the status quo, to daily life. First, we provided a comprehensive survey of the research field of emotion recognition among couples (Paper 1). Second, we leveraged insights from psychology research and deep transfer learning approaches to develop machine learning systems to recognize each partner’s emotions using lab data from Dutch-speaking couples in Belgium (Paper 2) and German-speaking couples in Switzerland (Papers 3 and 4). We also performed emotion recognition using data from German-speaking elderly individuals (not romantic partners) in Germany (Paper 5) given the target use case for our emotion recognition system consisted of partners who were elderly and spoke German. Third, we developed ubiquitous smartwatch and smartphone systems — VADLite and DyMand — to collect relevant multimodal sensor data and self-report emotion data from the daily life interactions of German-speaking, Swiss-based couples managing type 2 diabetes (Papers 6 and 7). Finally, we developed and evaluated machine learning systems for recognizing each partner’s emotions using the collected multimodal real-world smartwatch data  — heart rate, accelerometer, gyroscope, and speech (Paper 8).

This thesis contributes toward building automated emotion recognition systems that would eventually enable partners to monitor their emotions in daily life and enable the delivery of interventions to improve their emotional well-being.

\clearpage
\section*{Résumé}
\addcontentsline{toc}{section}{Résumé}
Les couples gèrent généralement ensemble les maladies chroniques et la gestion a un impact émotionnel sur les patients et leurs partenaires amoureux. Par conséquent, la reconnaissance des émotions de chaque partenaire dans la vie quotidienne pourrait donner un aperçu de leur bien-être émotionnel dans la gestion des maladies chroniques. Les émotions des partenaires sont actuellement déduites en laboratoire et dans la vie quotidienne à l'aide d'auto-rapports qui ne sont pas pratiques pour l'évaluation continue des émotions ou de rapports d'observateurs qui sont manuels, chronophages et coûteux. Actuellement, il n'existe pas de synthèse exhaustive des travaux sur la reconnaissance des émotions chez les couples. De plus, les approches de reconnaissance des émotions chez les couples se sont (1) concentrées sur les couples anglophones aux États-Unis, (2) ont utilisé des données recueillies auprès du laboratoire et (3) ont effectué une reconnaissance en utilisant les évaluations des observateurs plutôt que les émotions autodéclarées / subjectives du partenaire. .

Dans cet ensemble de travaux contenus dans cette thèse (8 articles — 5 publiés et 3 actuellement en cours de révision dans diverses revues), nous comblons le vide bibliographique actuel sur la reconnaissance des émotions des couples, développons des systèmes de reconnaissance des émotions en utilisant 161 heures de données provenant d'un total de 1,051 personnes et contribuent à faire passer la reconnaissance des émotions des couples du laboratoire, qui est le statu quo, à la vie quotidienne. Tout d'abord, nous avons fourni une enquête complète sur le domaine de recherche de la reconnaissance des émotions chez les couples (article 1). Deuxièmement, nous avons tiré parti des connaissances de la recherche en psychologie et des approches “deep transfer learning” pour développer des systèmes d'apprentissage automatique permettant de reconnaître les émotions de chaque partenaire à l'aide de données de laboratoire de couples néerlandophones en Belgique (article 2) et de couples germanophones en Suisse (articles 3 et 4). Nous avons également effectué la reconnaissance des émotions à l'aide de données provenant de personnes âgées germanophones (pas de partenaires romantiques) en Allemagne (article 5), étant donné que le cas d'utilisation cible de notre système de reconnaissance des émotions était composé de partenaires âgés et parlant allemand. Troisièmement, nous avons développé des systèmes ubiquitaires de montres intelligentes et de smartphones — VADLite et DyMand — pour collecter des données de capteurs multimodaux pertinentes et auto-déclarer des données émotionnelles à partir des interactions quotidiennes de couples germanophones basés en Suisse qui gèrent le diabète de type 2 (articles 6 et 7) . Enfin, nous avons développé et évalué des systèmes d'apprentissage automatique pour reconnaître les émotions de chaque partenaire à l'aide des données de smartwatch multimodales du monde réel collectées - fréquence cardiaque, accéléromètre, gyroscope et parole (article 8).

Cette thèse contribue à la construction de systèmes automatisés de reconnaissance des émotions qui permettraient éventuellement aux partenaires de surveiller leurs émotions dans la vie quotidienne et de permettre la réalisation d'interventions pour améliorer leur bien-être émotionnel.

\clearpage
\section*{Acknowledgement}
\addcontentsline{toc}{section}{Acknowledgement}
First, I would like to express my gratitude to Prof. Dr. Elgar Fleisch for the opportunity to do my PhD in the Center for Digital Health Interventions. The ethos of doing research that is both technically rigorous and societally relevant will always stick with me. The continuous asking of what our technical results mean in the real-world has influenced my drive to keep working on answering research questions that make a difference in the lives of people. I thank Prof. Dr.  Tobias Kowatch for his ever-available support, feedback and in particular, his infectious enthusiasm and optimism about the projects in our center.

I’m grateful to Prabhakaran Santhanam, my co-laborer in the DyMand project for all his assistance and support. I thank my DyMand project collaborators at the University of Zurich for the opportunity to work together: Professor Urte Scholz, Professor Guy Bodenmann, Dr. Janina Luscher and Dr. Theresa Pauly.

I thank various postdocs and professors for their feedback on my work at various stages of my PhD: Prof. Cecilia Mascolo, Prof. David Kotz, Prof. Dr. Elliott Ash, Prof. Emily Provost, Prof. Dr. Felix Wortmann, Prof. Dr. Florian Wangenheim, Prof. Dr. Peter Hilpert, Prof. Dr. Petra Schmid, Dr. Laura Sels, Prof. Dr.  Stefan  Feuerriege, Prof. Temiloluwa Prioleau, Prof. Dr. Verena Tiefenbeck  

I’m grateful to several master’s students I’ve been privileged to work with who contributed to these works: Madhav Sachdeva, Malgorzata Speichert, Marine Hoche, Jacopo Biggiogera, Xiangyu Zhao. I thank the following research assistants for assisting with data collection, coding, annotating and transcribing our dataset: Kemeng Zhang, Kwabena Atobra, Elena Luzi, Denis Adamec, and Luljeta Isaki.

I’m grateful to all the individuals who gave feedback on my work in conferences, research visits, or paper reviews and in particular, my colleagues in our research group for the feedback and contributions along these years: Dr. Anselma Wörner, Caterina Bérubé, Davide Clares, Dr. Dominik Rüegger, Eva van Weenen, Dr. Filipe Barata, Fabian Schneider, Dr. Florian Künzler, Gisbert Teepe, Dr. Iris Shih, Dr. Jan-Niklas Kramer, Kevin Koch, Dr. Klaus Fuchs, Dr. Liliane Ableitner, Mara Nägelin, Martin Maritsch, Dr. Peter Tinschert, Raphael Weibel, Robert Jakob, Simon Föll, Dr. Shu Liu, Yanick Lukic. 

I thank Monica Heinz, Elisabeth Keller, Judith Holzheimer, Olivia Keller for assisting with the administrative aspects of doing PhD at ETH.

I thank my family and friends for their support throughout this journey.

Last but not least, I thank God for His grace which has been sufficient through this journey, and the strength to trust that His plans and purpose for me are always good, through it all. To Him be the glory.

\clearpage
\tableofcontents
\addcontentsline{toc}{section}{Contents}
\clearpage

\pagenumbering{arabic}
\include{intro}

\clearpage
\printbibliography 
\addcontentsline{toc}{section}{References}

\clearpage
\section*{Paper 1: Survey of Couples' Emotion Recognition}
\addcontentsline{toc}{section}{Paper 1:  Survey of Couples' Emotion Recognition}

\clearpage
\includepdf[pages=-]{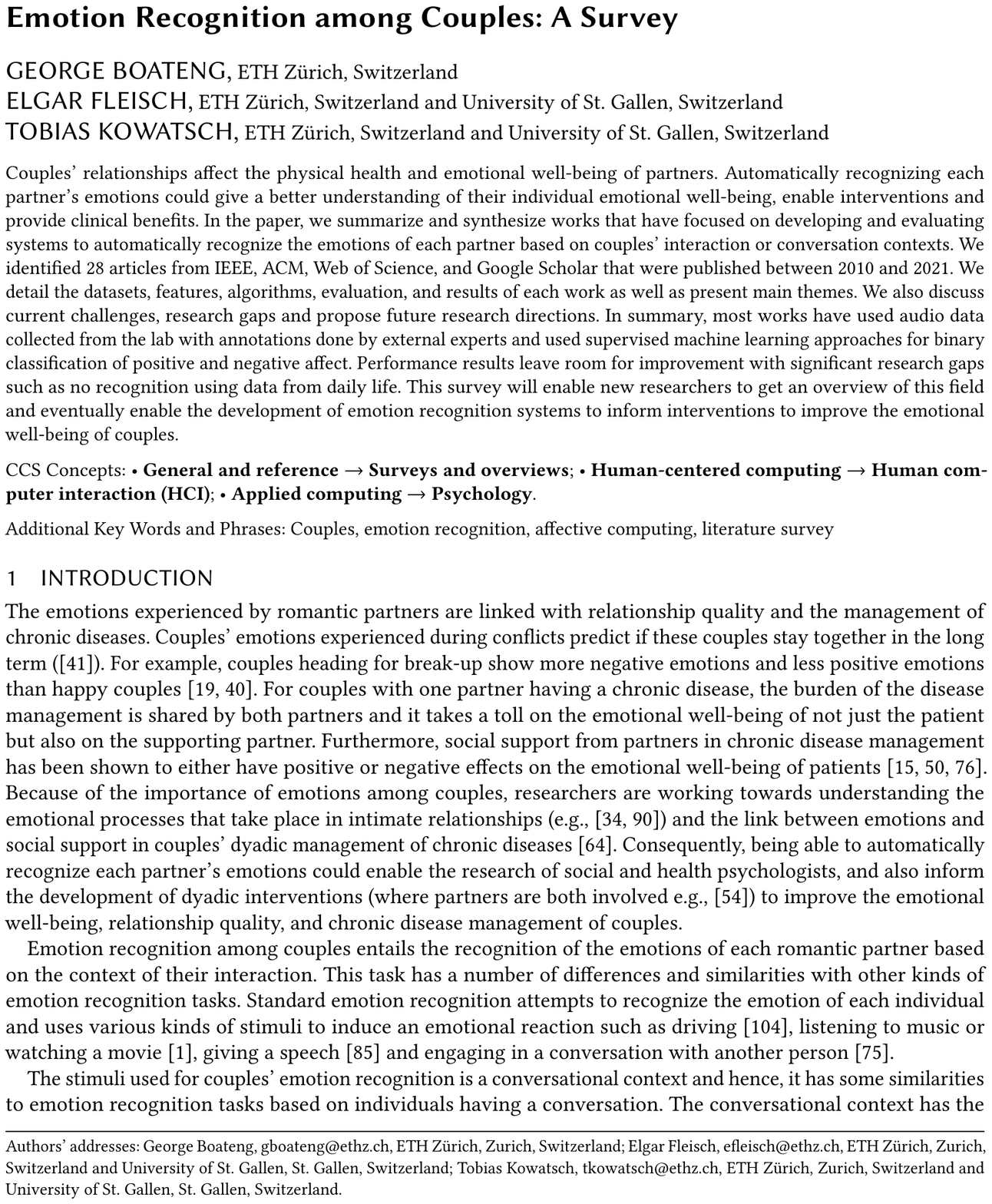}

\clearpage
\section*{Paper 2: Peak-End Rule}
\addcontentsline{toc}{section}{Paper 2: Peak-End Rule}

\clearpage
\includepdf[pages=-]{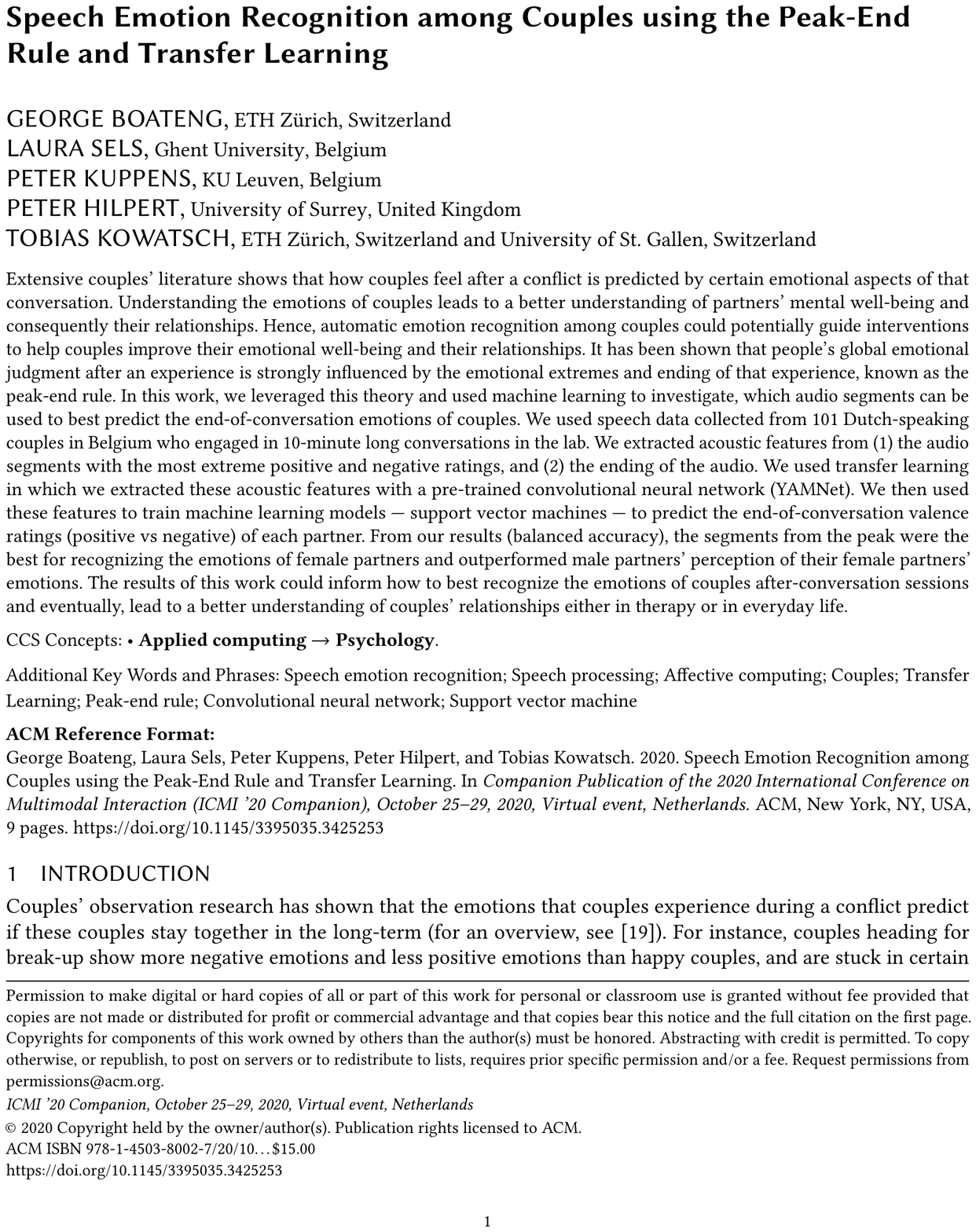}

\clearpage
\section*{Paper 3: BERT meets LIWC}
\addcontentsline{toc}{section}{Paper 3: BERT meets LIWC}

\clearpage
\includepdf[pages=-]{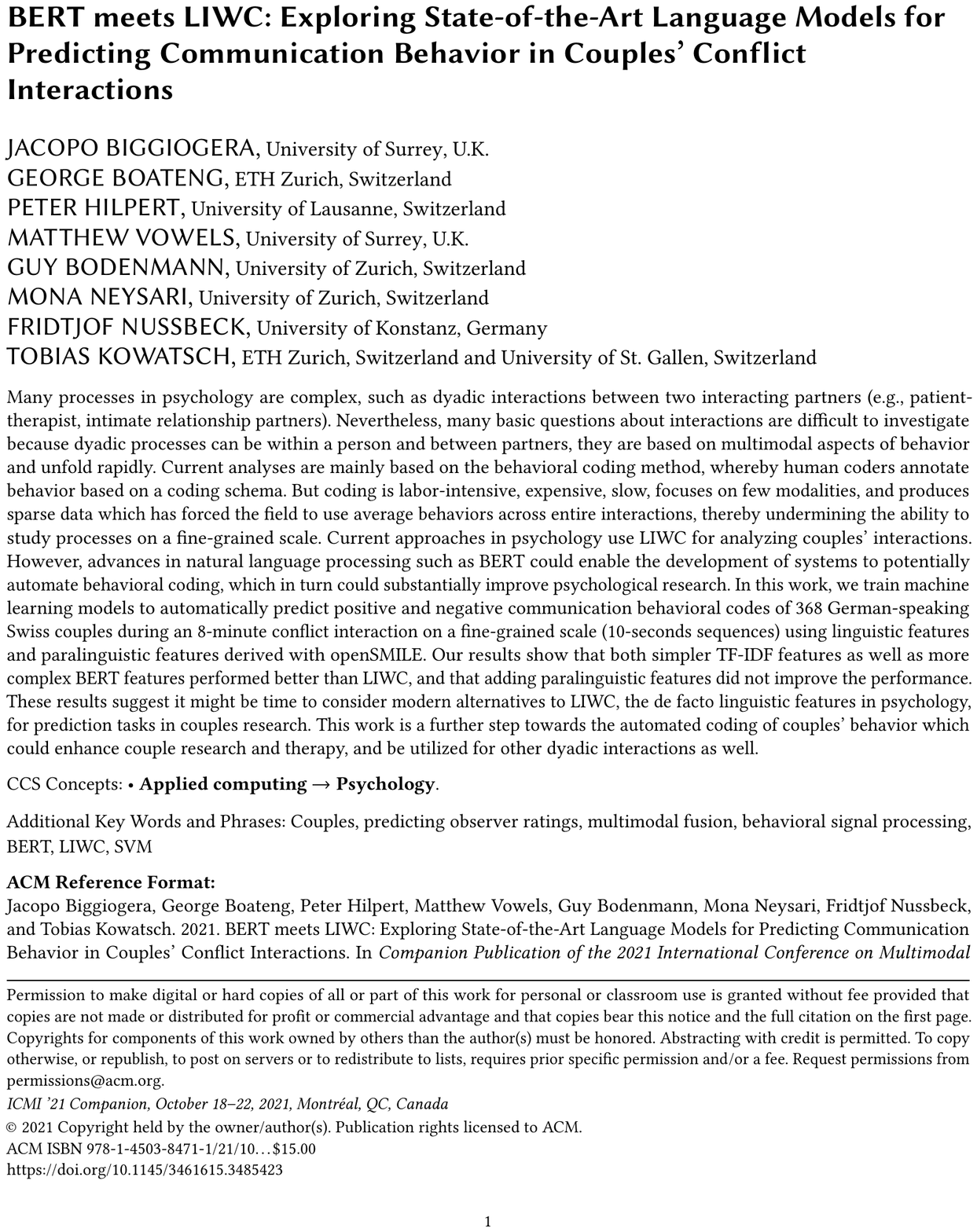}

\clearpage
\section*{Paper 4: You made me feel this way}
\addcontentsline{toc}{section}{Paper 4: You made me feel this way}

\clearpage
\includepdf[pages=-]{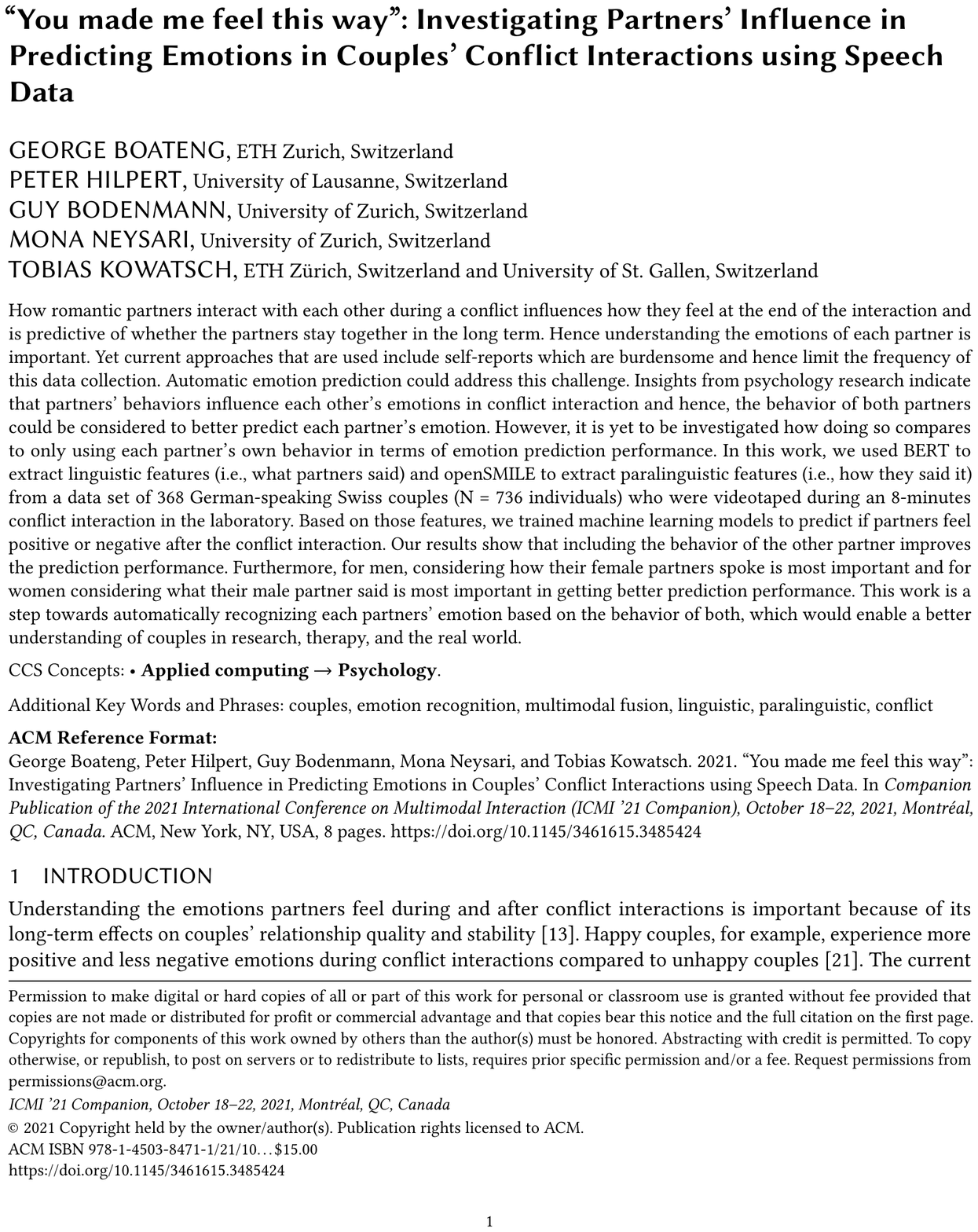}

\clearpage
\section*{Paper 5: Elderly Emotion Recognition}
\addcontentsline{toc}{section}{Paper 5: Elderly Emotion Recognition}

\clearpage
\includepdf[pages=-]{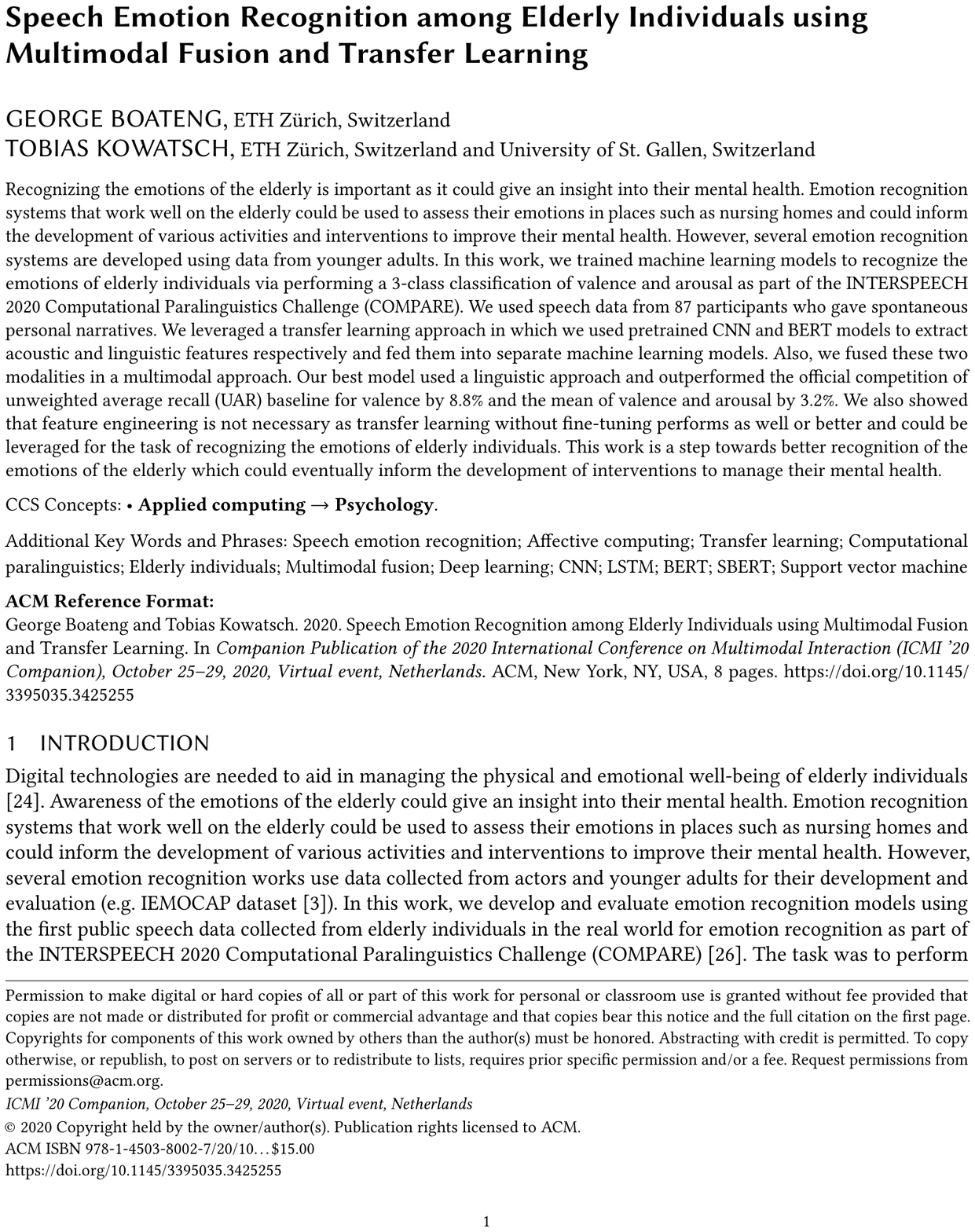}

\clearpage
\section*{Paper 6: VADLite}
\addcontentsline{toc}{section}{Paper 6: VADLite}

\clearpage
\includepdf[pages=-]{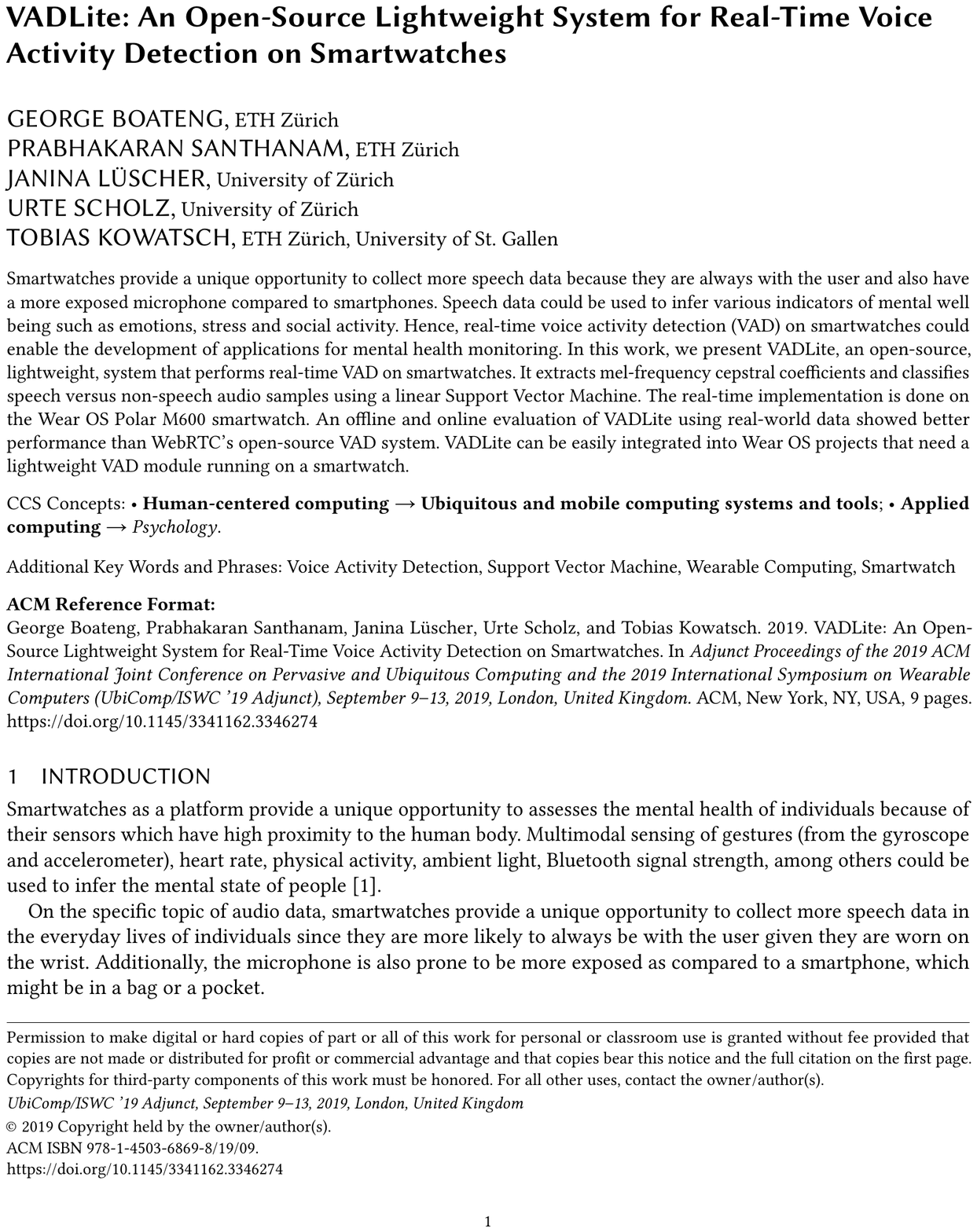}

\clearpage
\section*{Paper 7: DyMand}
\addcontentsline{toc}{section}{Paper 7: DyMand}

\clearpage
\includepdf[pages=-]{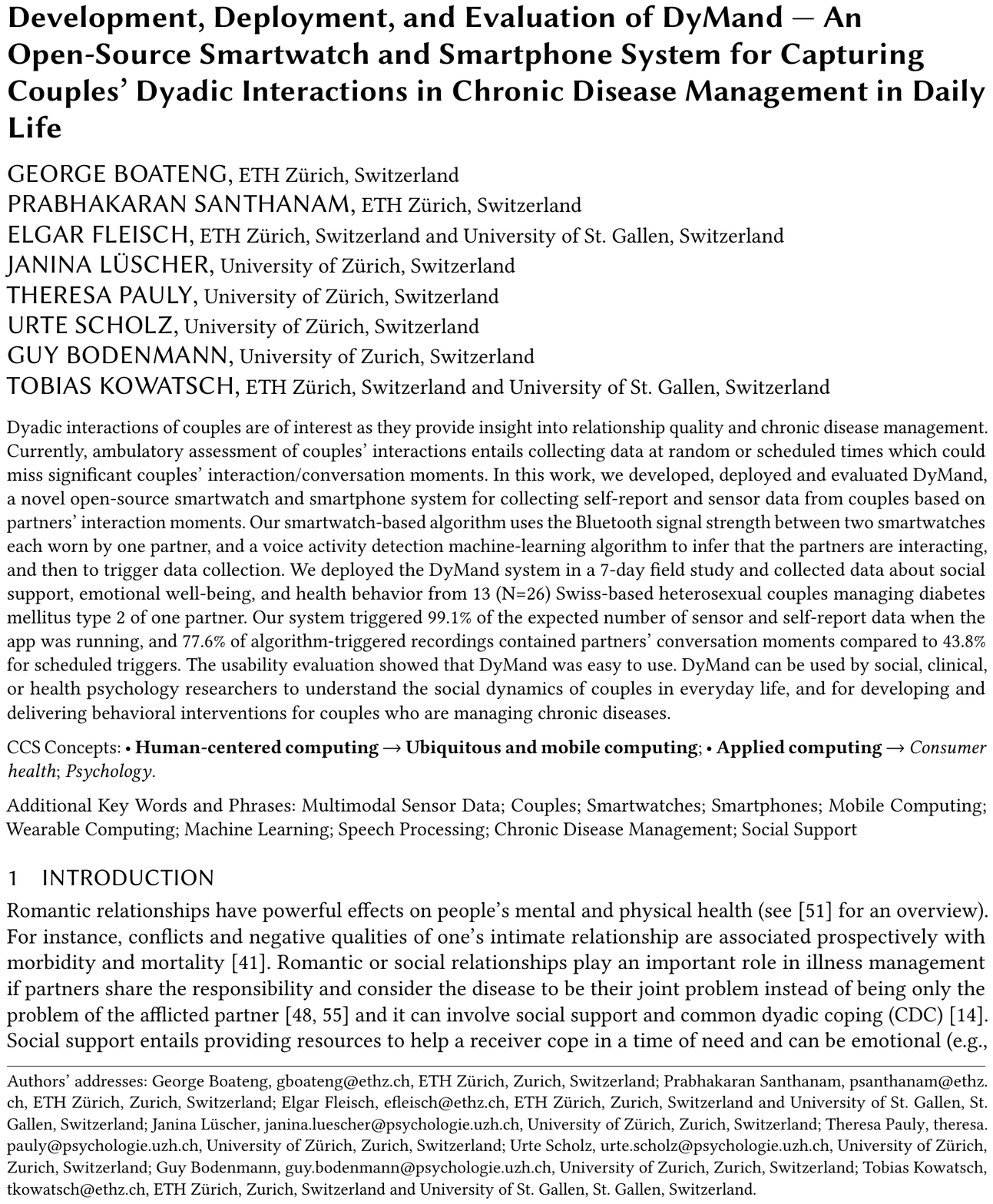}

\clearpage
\section*{Paper 8: Are you ok, honey?}
\addcontentsline{toc}{section}{Paper 8: Are you ok, honey?}

\clearpage
\includepdf[pages=-]{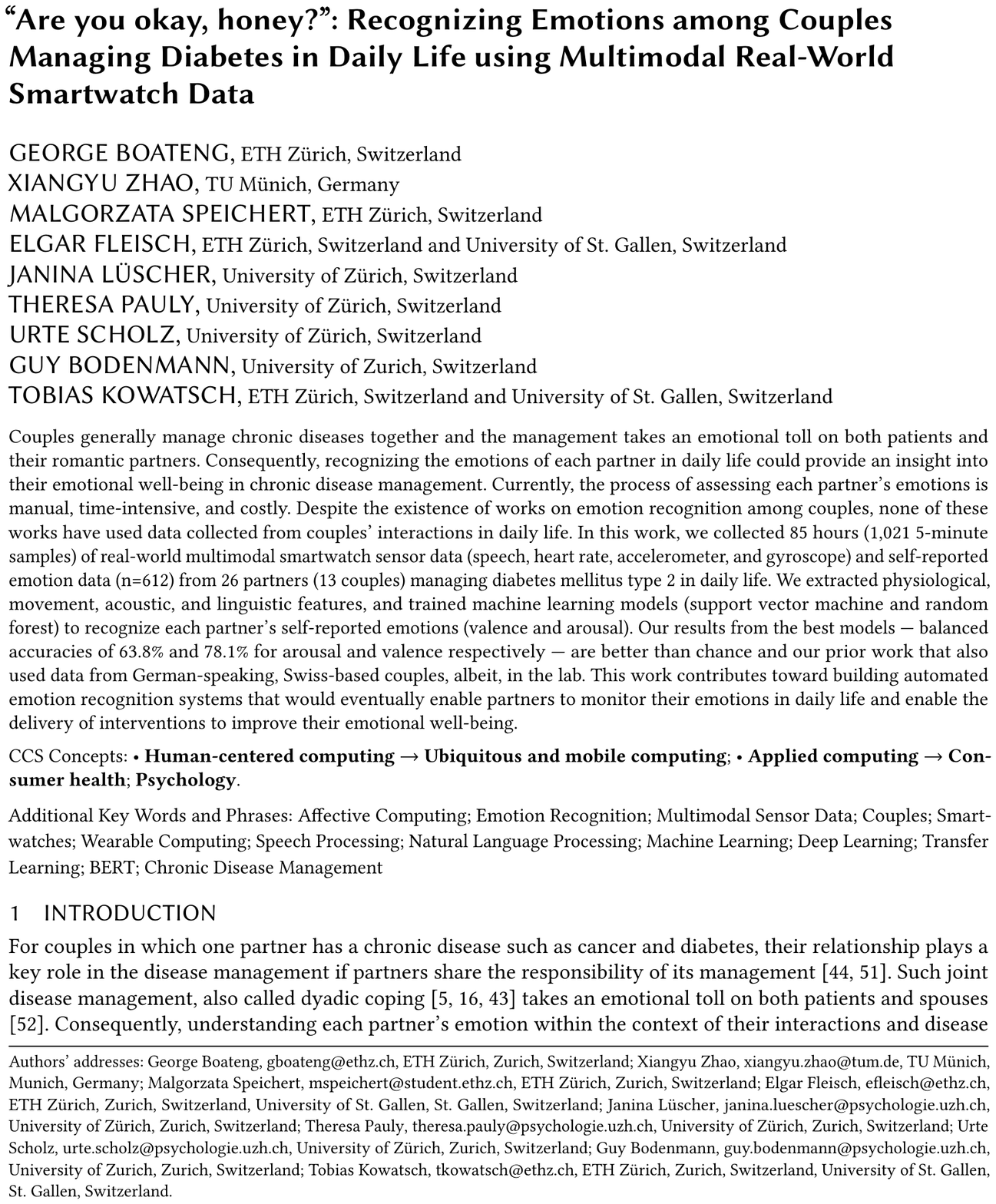}





\end{document}

%% file: intro.tex
\section*{Introduction}
\addcontentsline{toc}{section}{Introduction}
Romantic partners’ emotions have links with the quality of their relationship, and disease management when one partner has a chronic disease. Couples heading for break-up show more negative emotions and less positive emotions than happy couples \cite{carstensen1995, gottman2005}. And for couples in which one partner has a chronic disease such as cancer and diabetes, its joint management called dyadic coping \cite{badr2017, bodenmann1997, revenson2011} takes an emotional toll on both patients and spouses  \cite{settineri2014}.  Consequently, recognizing the emotions of each partner could unobtrusively provide an insight into the quality of their relationship, and the emotional well-being of each partner, and enable the triggering of interventions to improve their relationship and chronic disease management.

There are mainly two models of emotions: categorical and dimensional. Categorical emotions are based on the six basic emotions proposed by Ekman: happiness, sadness, fear, anger, disgust, and surprise  \cite{ekman1997}. Dimensional approaches mainly use two dimensions: valence (pleasure) and arousal which are based on Russell’s circumplex model of emotions \cite{russell1980}. Valence refers to how negative to positive a person feels and arousal refers to how sleepy to active a person feels. Using these two dimensions, several categorical emotions can be placed and grouped into the four quadrants: high arousal and negative valence (e.g., angry), low arousal and negative valence (e.g., depressed), low arousal and positive valence (e.g., relaxed), and high arousal and positive valence (e.g., excited) \cite{russell1980}.

Currently, the emotions of romantic partners are inferred in the lab and in daily life using two approaches: self-report and observer reports \cite{boateng2022a}.  In the lab, couples are asked to have an emotionally charged conversation that is videotaped. Afterward, each partner provides emotion ratings, for example, by adjusting a joystick while watching the videos \cite{roberts2007} or by using a validated affect instrument such as the Multidimensional Mood questionnaire \cite{steyer1997}. Also, people are trained to watch the video recordings and use a coding scheme (e.g., SPAFF \cite{coan2007}) to rate the emotional behavior of each partner. In the case of daily life, couples are periodically asked to complete self-reports \cite{schoebi2008} such as the PANAS \cite{watson1988} or observers can code audio data collected from couples’ daily life interactions \cite{robbins2014}. Collecting self-reports in daily life are obtrusive and impractical for continuous emotion assessment (e.g., every minute). Furthermore, self-reports could be biased (for example, if the partner desires to project a certain emotion rating rather than how they really feel) and may not reflect the partner’s true emotion. The manual coding process with observer reports is costly and time-consuming as multiple coders need to be trained for this task \cite{kerig2004} and suffers from inter-rater reliability issues \cite{heyman2001, metallinou2013}. Additionally, observer reports do not reflect the subjective emotions of partners, but rather, an external person’s assessment based on expressed behavior. Automated recognition of each partner’s emotion could potentially address these limitations by leveraging sensor data (e.g., audio).

Emotion recognition among couples is the task of automatically recognizing the emotions of romantic partners based on their conversation or interaction context \cite{boateng2022a}. In particular, it entails recognizing each partner’s emotions for every utterance/speaker turn or every few seconds — local emotion — or for the whole conversation — global emotion. These emotion ratings are the self-reports provided by the partners or observer ratings provided by external individuals. This task differs from other kinds of emotion recognition tasks mainly by the kind of stimuli that induces emotions. Some stimuli are driving  \cite{zepf2020}, listening to music or watching a movie \cite{abadi2015}, and conversation between people  \cite{poria2019}. Couples’ emotion recognition is similar to emotion recognition tasks whose stimuli are conversations since it uses a conversational context. However, its uniqueness lies in the fact that the two interacting individuals are in a romantic relationship. Consequently, various insights from psychology about couples’ interaction dynamics could be leveraged to recognize each partner’s emotions. For example, romantic partners influence each other when interacting, and that insight has been used for couples’ emotion recognition (e.g.,  \cite{chakravarthula2018, boateng2021}). Currently, there exists no comprehensive overview of works on emotion recognition among couples. Furthermore, approaches for emotion recognition among couples have (1) focused on English-speaking couples in the U.S., (2) used data collected from the lab, and (3) performed recognition using observer ratings rather than partner’s self-reported / subjective emotions. 

In this body of work contained in this thesis (8 papers), we fill the current literature gap on couples’ emotion recognition, develop emotion recognition systems using 161 hours of data from a total of 1,051 individuals, and make contributions towards taking couples’ emotion recognition from the lab which is the status quo, to daily life. First, we provided a comprehensive survey of the research field of emotion recognition among couples (Paper 1 \cite{boateng2022a}). Second, we leveraged insights from psychology research and deep transfer learning approaches to develop machine learning systems to recognize each partner’s emotions using lab data from Dutch-speaking couples in Belgium (Paper 2 \cite{boateng2020a}) and German-speaking couples in Switzerland (Paper 3 \cite{biggiogera2021} and Paper 4 \cite{boateng2021}). We also performed emotion recognition using data from German-speaking elderly individuals (not romantic partners) in Germany (Paper 5 \cite{boateng2020b}) given the target use case for our emotion recognition system consisted of partners who were elderly and spoke German. Third, we developed ubiquitous smartwatch and smartphone systems — VADLite and DyMand — to collect relevant multimodal sensor data and self-report emotion data from the daily life interactions of German-speaking, Swiss-based couples managing type 2 diabetes (Paper 6 \cite{boateng2019a} and Paper 7 \cite{boateng2022b}). Finally, we developed and evaluated machine learning systems for recognizing each partner’s emotions using the collected multimodal real-world smartwatch data  — heart rate, accelerometer, gyroscope, and speech (Paper 8 \cite{boateng2022c}). Of the 8 papers included in this thesis, 5 have been published and 3 are currently under review in various journals. The rest of the thesis contains an overview of the 8 papers with the reference of each paper provided, an abstract of the papers, their scientific contributions, my specific contribution to each of the works, and the full text of all 8 papers.

\subsection*{Paper 1: Survey of Couples’ Emotion Recognition}
\subsubsection*{Reference}
\textbf{George Boateng}, Elgar Fleisch, Tobias Kowatch. “Emotion Recognition among Couples: A Survey”. \textit{ACM Computing Surveys \textbf{(under review)}}. \url{https://arxiv.org/abs/2202.08430} 

\subsubsection*{Abstract}
Couples’ relationships affect the physical health and emotional well-being of partners. Automatically recognizing each partner’s emotions could give a better understanding of their individual emotional well-being, enable interventions and provide clinical benefits. In the paper, we summarize and synthesize works that have focused on developing and evaluating systems to automatically recognize the emotions of each partner based on couples’ interaction or conversation contexts. We identified 28 articles from IEEE, ACM, Web of Science, and Google Scholar that were published between 2010 and 2021. We detail the datasets, features, algorithms, evaluation, and results of each work as well as present main themes. We also discuss current challenges, research gaps and propose future research directions. In summary, most works have used audio data collected from the lab with annotations done by external experts and used supervised machine learning approaches for binary classification of positive and negative affect. Performance results leave room for improvement with significant research gaps such as no recognition using data from daily life. This survey will enable new researchers to get an overview of this field and eventually enable the development of emotion recognition systems to inform interventions to improve the emotional well-being of couples.

\subsubsection*{Contribution}
Our contributions are as follows: (1) the first survey and comprehensive overview of the field of emotion recognition among couples which would enable new researchers to get started in the field easily (2) background on emotion models, elicitation, and annotation approaches (3) details of the datasets, features, modalities, algorithms, evaluation, and results of each work (4) modeling approaches that consider the unique context of couples’ interactions (5) discussion of current challenges, research gaps and proposal of future research directions. For this paper, I performed the search of key terms in various databases, selected papers based on the inclusion and exclusion criteria, read and wrote summaries of the selected papers, wrote the first draft of this paper, and then the final draft after receiving feedback.

\subsection*{Paper 2: Peak-End Rule}
\subsubsection*{Reference}
\textbf{George Boateng}, Laura Sels, Peter Kuppens, Peter Hilpert, Urte Scholz, and Tobias Kowatsch. 2020. “Speech Emotion Recognition among Couples using the Peak-End Rule and Transfer Learning”. In \textit{Companion Publication of the 2020 International Conference on Multimodal Interaction (ICMI ’20 Companion), October 25–29, 2020, Virtual event, Netherlands}. ACM, New York, NY, USA, 5 pages. \url{https://doi.org/10.1145/3395035.3425253} 

\subsubsection*{Abstract}
Extensive couples’ literature shows that how couples feel after a conflict is predicted by certain emotional aspects of that conversation. Understanding the emotions of couples leads to a better understanding of partners’ mental well-being and consequently their relationships. Hence, automatic emotion recognition among couples could potentially guide interventions to help couples improve their emotional well-being and their relationships. It has been shown that people’s global emotional judgment after an experience is strongly influenced by the emotional extremes and ending of that experience, known as the peak-end rule. In this work, we leveraged this theory and used machine learning to investigate which audio segments can be used to best predict the end-of-conversation emotions of couples. We used speech data collected from 101 Dutch-speaking couples in Belgium who engaged in 10-minute long conversations in the lab. We extracted acoustic features from (1) the audio segments with the most extreme positive and negative ratings — the peak, and (2) the ending of the audio — the end. We used transfer learning in which we extracted these acoustic features with a pre-trained convolutional neural network (YAMNet). We then used these features to train machine learning models — support vector machines — to predict the end-of-conversation valence ratings (positive vs negative) of each partner. From our results (balanced accuracy), the segments from the peak were the best for recognizing the emotions of female partners and outperformed male partners’ perception of their female partners’ emotions. The results of this work could inform how to best recognize the emotions of couples after-conversation sessions and eventually, lead to a better understanding of couples’ relationships either in therapy or in everyday life.

\subsubsection*{Contribution}
Our contributions are as follows: (1) exploration of the best way to recognize the emotions of couples after a conversation (5 - 10 minutes) through the peak-end rule lens using deep transfer learning — classification of end-of-conversation valence using acoustic features from the emotional peaks and end of the audio  (2) use of a unique dataset — real-world data collected from Dutch- speaking couples with self-ratings of emotions (3) proposal and computation of a “partner perception baseline” for emotion recognition within the context of couples interactions that leverage each partner’s perception of their partner’s emotions and gives an estimate of how well each partner could infer their partner's emotion. For this paper, I conceptualized the key idea, preprocessed the dataset, extracted features, implemented the machine learning experiments, wrote the first draft of the paper, and then the final draft after receiving feedback.

\subsection*{Paper 3: BERT meets LIWC}
\subsubsection*{Reference}
Jacopo Biggiogera, \textbf{George Boateng}, Peter Hilpert, Matthew Vowels, Guy Bodenmann, Mona Neysari, Fridtjof Nussbeck, Tobias Kowatsch. “BERT meets LIWC: Exploring State-of-the-Art Language Models for Predicting Communication Behavior in Couples’ Conflict Interactions”. In \textit{Companion Publication of the 2021 International Conference on Multimodal Interaction (ICMI ’21 Companion), October 18–22, 2021, Montréal, QC, Canada}. ACM, New York, NY, USA, 5 pages. \url{ https://doi.org/10.1145/3461615.3485423} 

\subsubsection*{Abstract}
Many processes in psychology are complex, such as dyadic interactions between two interacting partners (e.g., patient-therapist, intimate relationship partners). Nevertheless, many basic questions about interactions are difficult to investigate because dyadic processes can be within a person and between partners, they are based on multimodal aspects of behavior and unfold rapidly. Current analyses are mainly based on the behavioral coding method, whereby human coders annotate behavior based on a coding schema. But coding is labor-intensive, expensive, slow, focuses on few modalities, and produces sparse data which has forced the field to use average behaviors across entire interactions, thereby undermining the ability to study processes on a fine-grained scale. Current approaches in psychology use LIWC for analyzing couples’ interactions. However, advances in natural language processing such as BERT could enable the development of systems to potentially automate behavioral coding, which in turn could substantially improve psychological research. In this work, we train machine learning models to automatically predict positive and negative communication behavioral codes of 368 German-speaking Swiss couples during an 8-minute conflict interaction on a fine-grained scale (10-seconds sequences) using linguistic features and paralinguistic features derived with openSMILE. Our results show that both simpler TF-IDF features as well as more complex BERT features performed better than LIWC, and that adding paralinguistic features did not improve the performance. These results suggest it might be time to consider modern alternatives to LIWC, the de facto linguistic features in psychology, for prediction tasks in couples research. This work is a further step towards the automated coding of couples’ behavior which could enhance couple research and therapy, and be utilized for other dyadic interactions as well.

\subsubsection*{Contribution}
Our contributions are as follows: (1) an evaluation of the predictive capability of BERT vis-à-vis LIWC in the context of the automatic recognition of couples’ communication behavioral codes on a fine-grained time scale (every 10 seconds) (2) an investigation into how the addition of paralinguistic features affects prediction performance (3) the use of a unique dataset — spontaneous, naturalistic, speech data collected from German-speaking Swiss couples (n=368 couples, N=736 participants), and the largest ever such dataset used in the literature for automatic coding of couples’ behavior. For this paper, I co-conceptualized the key idea, extracted paralinguistic and linguistic (BERT) features, provided feedback on the machine learning experiments, and co-wrote the first and final drafts of the paper.

\subsection*{Paper 3: You made me feel this way}
\subsubsection*{Reference}
\textbf{George Boateng}, Peter Hilpert, Guy Bodenmann, Mona Neysari, Tobias Kowatsch. ““You made me feel this way”: Investigating Partners’ Influence in Predicting Emotions in Couples’ Conflict Interactions using Speech Data”. In \textit{Companion Publication of the 2021 International Conference on Multimodal Interaction (ICMI ’21 Companion), October 18–22, 2021, Montréal, QC, Canada}. ACM, New York, NY, USA, 5 pages. \url{ https://doi.org/10.1145/3461615.3485424} 

\subsubsection*{Abstract}
How romantic partners interact with each other during a conflict influences how they feel at the end of the interaction and is predictive of whether the partners stay together in the long term. Hence understanding the emotions of each partner is important. Yet current approaches that are used include self-reports which are burdensome and hence limit the frequency of this data collection. Automatic emotion prediction could address this challenge. Insights from psychology research indicate that partners’ behaviors influence each other’s emotions in conflict interaction and hence, the behavior of both partners could be considered to better predict each partner’s emotion. However, it is yet to be investigated how doing so compares to only using each partner’s own behavior in terms of emotion prediction performance. In this work, we used BERT to extract linguistic features (i.e., what partners said) and openSMILE to extract paralinguistic features (i.e., how they said it) from a data set of 368 German-speaking Swiss couples (N = 736 individuals) who were videotaped during an 8-minutes conflict interaction in the laboratory. Based on those features, we trained machine learning models to predict if partners feel positive or negative after the conflict interaction.  Our results show that including the behavior of the other partner improves the prediction performance. Furthermore, for men, considering how their female partners spoke is most important and for women considering what their male partner said is most important in getting better prediction performance. This work is a step towards automatically recognizing each partner’s emotion based on the behavior of both, which would enable a better understanding of couples in research, therapy, and the real world.

\subsubsection*{Contribution}
This work builds upon Paper 2 by using (1) the same data set, (2) global emotion ratings rather than local emotion ratings, (3) self-reports rather than observer reports, and (4) BERT which was shown to perform well for linguistic feature extraction in Paper 2. Our contributions are as follows: (1) an evaluation of how well a partner’s own linguistic and paralinguistic features predict one’s own end-of-conversation emotion (2) an investigation of how the prediction performance changes when including one’s partner’s features (linguistic, paralinguistic, and both) (3) the use of a unique dataset — spontaneous, naturalistic, speech data collected from German-speaking, Swiss couples (n=368 couples, N=736 participants), which is the largest ever such dataset used in the literature for automatic recognition of partners’ end-of-conversation emotion. For this paper, I co-conceptualized the key idea, preprocessed the dataset, extracted features, implemented the machine learning experiments, and co-wrote the first and final drafts of the paper.

\subsection*{Paper 5: Elderly Emotion Recognition}
\subsubsection*{Reference}
\textbf{George Boateng} and Tobias Kowatsch. 2020. “Speech Emotion Recognition among Elderly Individuals using Multimodal Fusion and Transfer Learning”. In \textit{Companion Publication of the 2020 International Conference on Multimodal Interaction (ICMI ’20 Companion), October 25–29, 2020, Virtual event, Netherlands}. ACM, New York, NY, USA, 5 pages. \url{https://doi.org/10.1145/3395035.3425255}

\subsubsection*{Abstract}
Recognizing the emotions of the elderly is important as it could give an insight into their mental health. Emotion recognition systems that work well on the elderly could be used to assess their emotions in places such as nursing homes and could inform the development of various activities and interventions to improve their mental health. However, several emotion recognition systems are developed using data from younger adults. In this work, we trained machine learning models to recognize the emotions of elderly individuals via performing a 3-class classification of valence and arousal as part of the INTERSPEECH 2020 Computational Paralinguistics Challenge (COMPARE). We used speech data from 87 participants who gave spontaneous personal narratives. We leveraged a transfer learning approach in which we used pretrained CNN and BERT models to extract acoustic and linguistic features respectively and fed them into separate machine learning models. Also, we fused these two modalities in a multimodal approach. Our best model used a linguistic approach and outperformed the official competition of unweighted average recall (UAR) baseline for valence by 8.8\% and the mean of valence and arousal by 3.2\%. We also showed that feature engineering is not necessary as transfer learning without fine-tuning performs as well or better and could be leveraged for the task of recognizing the emotions of elderly individuals. This work is a step towards better recognition of the emotions of the elderly which could eventually inform the development of interventions to manage their mental health.

\subsubsection*{Contribution}
Our contribution is the evaluation of transfer learning approaches to recognize the emotions of elderly individuals using a novel dataset — speech data collected from German-speaking elderly individuals via using a pretrained CNN model (YAMNET) to extract acoustic features and a pretrained Transformer language model — Bidirectional Encoder Representations from Transformers (BERT) to extract linguistic features. For this paper, I conceptualized the key idea, preprocessed the dataset, extracted features, implemented the machine learning experiments, wrote the first draft of the paper, and then the final draft after receiving feedback.

\subsection*{Paper 6: VADLite}
\subsubsection*{Reference}
\textbf{George Boateng}, Prabhakaran Santhanam, Janina L{\"u}scher, Urte Scholz, and Tobias Kowatsch. 2019. “VADLite: An Open-Source Lightweight System for Real-Time Voice Activity Detection on Smartwatches”. In \textit{Adjunct Proceedings of the 2019 ACM International Joint Conference on Pervasive and Ubiquitous Computing and the 2019 International Symposium on Wearable Computers (UbiComp/ISWC ’19 Adjunct), September 9–13, 2019, London, United Kingdom}. ACM, New York, NY, USA, 5 pages. \url{https://doi.org/10.1145/3341162.3346274}

\subsubsection*{Abstract}
Smartwatches provide a unique opportunity to collect more speech data because they are always with the user and also have a more exposed microphone compared to smartphones. Speech data could be used to infer various indicators of mental well-being such as emotions, stress, and social activity. Hence, real-time voice activity detection (VAD) on smartwatches could enable the development of applications for mental health monitoring. In this work, we present VADLite, an open-source, lightweight, system that performs real-time VAD on smartwatches. It extracts mel-frequency cepstral coefficients and classifies speech versus non-speech audio samples using a linear Support Vector Machine. The real-time implementation is done on the Wear OS Polar M600 smartwatch. An offline and online evaluation of VADLite using real-world data showed better performance than WebRTC's open-source VAD system. VADLite can be easily integrated into Wear OS projects that need a lightweight VAD module running on a smartwatch.

\subsubsection*{Contribution}
Our contribution is the development and evaluation of VADLite, an open-source lightweight software system \footnote{\url{https://https://bitbucket.org/Jojo29/vadlite/}} that performs real-time voice activity detection (VAD) on smartwatches and runs efficiently on constrained systems via using a linear support vector machine, consequently addressing the gap in obtaining an easy-to-use smartwatch VAD system. For this paper, I ran the user study to collect data, preprocessed the data, extracted features, implemented the machine learning experiments, designed and implemented the VADLite smartwatch app, wrote the first draft of the paper, and then the final draft after receiving feedback.

\subsection*{Paper 7: DyMand}
\subsubsection*{Reference}
\textbf{George Boateng}, Prabhakaran Santhanam, Elgar Fleisch, Janina L{\"u}scher, Theresa Pauly, Urte Scholz, and Tobias Kowatsch. “Development, Deployment, and Evaluation of DyMand--An Open-Source Smartwatch and Smartphone System for Capturing Couples' Dyadic Interactions in Chronic Disease Management in Daily Life”. \textit{Sensors \textbf{(under review)}}.  \url{https://arxiv.org/abs/2205.07671} 

\subsubsection*{Abstract}
Dyadic interactions of couples are of interest as they provide insight into relationship quality and chronic disease management. Currently, ambulatory assessment of couples’ interactions entails collecting data at random or scheduled times which could miss significant couples’ interaction/conversation moments. In this work, we developed, deployed, and evaluated DyMand, a novel open-source smartwatch and smartphone system for collecting self-report and sensor data from couples based on partners’ interaction moments. Our smartwatch-based algorithm uses the Bluetooth signal strength between two smartwatches each worn by one partner, and a voice activity detection machine-learning algorithm to infer that the partners are interacting, and then to trigger data collection. We deployed the DyMand system in a 7-day field study and collected data about social support, emotional well-being, and health behavior from 13 (N=26) Swiss-based heterosexual couples managing diabetes mellitus type 2 of one partner. Our system triggered 99.1\% of the expected number of sensor and self-report data when the app was running, and 77.6\% of algorithm-triggered recordings contained partners’ conversation moments compared to 43.8\% for scheduled triggers. The usability evaluation showed that DyMand was easy to use. DyMand can be used by social, clinical, or health psychology researchers to understand the social dynamics of couples in everyday life, and for developing and delivering behavioral interventions for couples who are managing chronic diseases.

\subsubsection*{Contribution}
This work incorporates VADLite from Paper 6. Our contributions are as follows: (1) designed and developed DyMand, a novel open-source smartwatch \footnote{\url{https://bitbucket.org/mobilecoach/dymandwatchclient/src/master/}}, and smartphone \footnote{\url{https://bitbucket.org/mobilecoach/dymand-mobilecoach-client/src/master/}} system that uses the Bluetooth signal strength between two smartwatches each worn by one partner, and a voice activity detection machine-learning algorithm to infer that the partners are interacting, and then to trigger sensor and self-report data collection (2) deployment and evaluation of DyMand in a field study with heterosexual couples in Switzerland that are managing type 2 diabetes (T2DM) of one partner. For this paper, I co-designed the DyMand system, implemented the DyMand smartwatch app, co-ran the user study to collect data, preprocessed the data, performed data analysis, wrote the first draft of the paper, and then the final draft after receiving feedback.

\subsection*{Paper 8: Are you ok, honey?}
\subsubsection*{Reference}
\textbf{George Boateng}, Xiangyu Zhao, Malgorzata Speichert, Elgar Fleisch, Janina L{\"u}scher, Theresa Pauly, Urte Scholz, Guy Bodenmann and Tobias Kowatsch. ““Are you okay, honey?”: Recognizing Emotions among Couples Managing Diabetes in Daily Life using Multimodal Smartwatch Data”. \textit{ACM Interact. Mob. Wearable Ubiquitous Technol \textbf{(under review)}}.   \url{http://arxiv.org/abs/2208.08909} 

\subsubsection*{Abstract}
Couples generally manage chronic diseases together and the management takes an emotional toll on both patients and their romantic partners. Consequently, recognizing the emotions of each partner in daily life could provide an insight into their emotional well-being in chronic disease management. Currently, the process of assessing each partner’s emotions is manual, time-intensive, and costly. Despite the existence of works on emotion recognition among couples, none of these works have used data collected from couples’ interactions in daily life.  In this work, we collected 85 hours (1,021 5-minute samples) of real-world multimodal smartwatch sensor data (speech, heart rate, accelerometer, and gyroscope) and self-reported emotion data (n=612) from 26 partners (13 couples) managing diabetes mellitus type 2 in daily life.  We extracted physiological, movement, acoustic, and linguistic features, and trained machine learning models (support vector machine and random forest) to recognize each partner’s self-reported emotions (valence and arousal). Our results from the best models — balanced accuracies of 63.8\% and 78.1\% for arousal and valence respectively — are better than chance and our prior work that also used data from German-speaking, Swiss-based couples, albeit, in the lab. This work contributes toward building automated emotion recognition systems that would eventually enable partners to monitor their emotions in daily life and enable the delivery of interventions to improve their emotional well-being.

\subsubsection*{Contributions}
This final paper builds upon all previous 7 papers. It summarizes and integrates the content of the survey paper (Paper 1), uses the data preprocessing, feature extraction and machine learning approaches detailed in Papers 2, 3, 4, and 5, and uses the smartwatch and smartphone systems from Papers 6 and 7 for data collection. This work is the first to recognize the emotions of romantic partners using data collected from everyday life. Our contributions are as follows (1) collection and use of a unique dataset — real-world, multimodal smartwatch sensor data from German-speaking, Swiss-based couples (N=13 couples, n=26 participants), which is the first such dataset used in the literature for automatic recognition of partners’ emotions (2) approaches for validating and quantifying data quality on manually coded, annotated and transcribed real-world speech data  (3) development and evaluation of a machine learning system to recognize the emotions of each partner using a wide variety of sensor data — acoustic, linguistic, heart rate, accelerometer, and gyroscope  (4) an investigation of the sensor modality combinations which result in the best emotion recognition performance of romantic partners. For this paper, I conceptualized the key idea, preprocessed the dataset, extracted features, implemented the machine learning experiments, wrote the first draft of the paper, and then the final draft after receiving feedback.

%% file: refs.bib
@article{abadi2015,
  title={DECAF: MEG-based multimodal database for decoding affective physiological responses},
  author={Abadi, Mojtaba Khomami and Subramanian, Ramanathan and Kia, Seyed Mostafa and Avesani, Paolo and Patras, Ioannis and Sebe, Nicu},
  journal={IEEE Transactions on Affective Computing},
  volume={6},
  number={3},
  pages={209--222},
  year={2015},
  publisher={IEEE}
}

@article{badr2017,
  title={Re-thinking dyadic coping in the context of chronic illness},
  author={Badr, Hoda and Acitelli, Linda K},
  journal={Current Opinion in Psychology},
  volume={13},
  pages={44--48},
  year={2017},
  publisher={Elsevier}
}

@inproceedings{boateng2019a,
  title={VADLite: an open-source lightweight system for real-time voice activity detection on smartwatches},
  author={Boateng, George and Santhanam, Prabhakaran and L{\"u}scher, Janina and Scholz, Urte and Kowatsch, Tobias},
  booktitle={Adjunct Proceedings of the 2019 ACM International Joint Conference on Pervasive and Ubiquitous Computing and Proceedings of the 2019 ACM International Symposium on Wearable Computers},
  pages={902--906},
  year={2019}
}

@inproceedings{boateng2020a,
  title={Speech Emotion Recognition among Couples using the Peak-End Rule and Transfer Learning},
  author={Boateng, George and Sels, Laura and Kuppens, Peter and Hilpert, Peter and Scholz, Urte and Kowatsch, Tobias},
  booktitle={Companion Publication of the 2020 International Conference on Multimodal Interaction (ICMI '20 Companion), October 25--29, 2020, Virtual event, Netherlands},
  year={2020}
}

@inproceedings{boateng2020b,
  title={Speech Emotion Recognition among Elderly Individuals using Multimodal Fusion and Transfer Learning},
  author={Boateng, George and Kowatsch, Tobias},
  booktitle={Companion Publication of the 2020 International Conference on Multimodal Interaction (ICMI '20 Companion), October 25--29, 2020, Virtual event, Netherlands},
  year={2020}
}

@inproceedings{boateng2021,
    author = {Boateng, George and Hilpert, Peter and Bodenmann, Guy and Neysari, Mona and Kowatsch, Tobias},
    title = {“You Made Me Feel This Way”: Investigating Partners’ Influence in Predicting Emotions in Couples’ Conflict Interactions Using Speech Data},
    year = {2021},
    isbn = {9781450384711},
    publisher = {Association for Computing Machinery},
    address = {New York, NY, USA},
    url = {https://doi.org/10.1145/3461615.3485424},
    doi = {10.1145/3461615.3485424},
    booktitle = {Companion Publication of the 2021 International Conference on Multimodal Interaction},
    pages = {390–394},
    numpages = {5},
    location = {Montreal, QC, Canada},
    series = {ICMI '21 Companion}
}

@article{boateng2022a,
  title={Emotion Recognition among Couples: A Survey},
  author={Boateng, George and Fleisch, Elgar and Kowatsch, Tobias},
  journal={arXiv preprint arXiv:2202.08430},
  year={2022}
}

@article{boateng2022b,
  title={Development, Deployment, and Evaluation of DyMand--An Open-Source Smartwatch and Smartphone System for Capturing Couples' Dyadic Interactions in Chronic Disease Management in Daily Life},
  author={Boateng, George and Santhanam, Prabhakaran and Fleisch, Elgar and L{\"u}scher, Janina and Pauly, Theresa and Scholz, Urte and Kowatsch, Tobias},
  journal={arXiv preprint arXiv:2205.07671},
  year={2022}
}

@misc{boateng2022c,
  doi = {10.48550/ARXIV.2208.08909},
  
  url = {https://arxiv.org/abs/2208.08909},
  
  author = {Boateng, George and Xiangyu Zhao and Malgorzata Speichert and Fleisch, Elgar and Lüscher, Janina and Pauly, Theresa and Scholz, Urte and Bodenmann, Guy and Kowatsch, Tobias},
  
  title = {"Are you okay, honey?": Recognizing Emotions among Couples Managing Diabetes in Daily Life using Multimodal Real-World Smartwatch Data},
  
  publisher = {arXiv},
  
  year = {2022},
  
  copyright = {Creative Commons Attribution Non Commercial No Derivatives 4.0 International}
}

@inbook{biggiogera2021,
    author = {Biggiogera, Jacopo and Boateng, George and Hilpert, Peter and Vowels, Matthew and Bodenmann, Guy and Neysari, Mona and Nussbeck, Fridtjof and Kowatsch, Tobias},
    title = {BERT Meets LIWC: Exploring State-of-the-Art Language Models for Predicting Communication Behavior in Couples’ Conflict Interactions},
    year = {2021},
    isbn = {9781450384711},
    publisher = {Association for Computing Machinery},
    address = {New York, NY, USA},
    url = {https://doi.org/10.1145/3461615.3485423},
    booktitle = {Companion Publication of the 2021 International Conference on Multimodal Interaction},
    pages = {385–389},
    numpages = {5}
}

@article{bodenmann1997,
  title={Dyadic coping-a systematic-transactional view of stress and coping among couples: Theory and empirical findings},
  author={Bodenmann, Guy},
  journal={European Review of Applied Psychology},
  volume={47},
  pages={137--140},
  year={1997},
  publisher={EDITIONS DU CENTRE DE PSYCHOLOGIE APPLIQUEE}
}

@article{carstensen1995,
  title={Emotional behavior in long-term marriage.},
  author={Carstensen, Laura L and Gottman, John M and Levenson, Robert W},
  journal={Psychology and aging},
  volume={10},
  number={1},
  pages={140},
  year={1995},
  publisher={American Psychological Association}
}

@article{chakravarthula2018,
  title={Modeling Interpersonal Influence of Verbal Behavior in Couples Therapy Dyadic Interactions},
  author={Chakravarthula, Sandeep Nallan and Baucom, Brian and Georgiou, Panayiotis},
  journal={arXiv preprint arXiv:1805.09436},
  year={2018}
}

@article{coan2007,
  title={The specific affect coding system (SPAFF)},
  author={Coan, James A and Gottman, John M},
  journal={Handbook of emotion elicitation and assessment},
  pages={267--285},
  year={2007}
}

@article{ekman1997,
  title={Universal facial expressions of emotion},
  author={Ekman, Paul and Keltner, Dacher},
  journal={Segerstrale U, P. Molnar P, eds. Nonverbal communication: Where nature meets culture},
  pages={27--46},
  year={1997}
}

@book{gottman2005,
  title={The mathematics of marriage: Dynamic nonlinear models},
  author={Gottman, John Mordechai},
  year={2005},
  publisher={MIT Press}
}

@article{heyman2001,
  title={Observation of couple conflicts: Clinical assessment applications, stubborn truths, and shaky foundations.},
  author={Heyman, Richard E},
  journal={Psychological assessment},
  volume={13},
  number={1},
  pages={5},
  year={2001},
  publisher={American Psychological Association}
}

@book{kerig2004,
  title={Couple observational coding systems},
  author={Kerig, Patricia K and Baucom, Donald H},
  year={2004},
  publisher={Taylor \& Francis}
}

@inproceedings{metallinou2013,
  title={Annotation and processing of continuous emotional attributes: Challenges and opportunities},
  author={Metallinou, Angeliki and Narayanan, Shrikanth},
  booktitle={2013 10th IEEE international conference and workshops on automatic face and gesture recognition (FG)},
  pages={1--8},
  year={2013},
  organization={IEEE}
}

@article{poria2019,
  title={Emotion recognition in conversation: Research challenges, datasets, and recent advances},
  author={Poria, Soujanya and Majumder, Navonil and Mihalcea, Rada and Hovy, Eduard},
  journal={IEEE Access},
  volume={7},
  pages={100943--100953},
  year={2019},
  publisher={IEEE}
}

@article{revenson2011,
  title={Couples coping with chronic illness.},
  author={Revenson, Tracey A and DeLongis, Anita},
  year={2011},
  publisher={Oxford University Press}
}

@article{roberts2007,
  title={Emotion elicitation using dyadic interaction tasks},
  author={Roberts, Nicole A and Tsai, Jeanne L and Coan, James A},
  journal={Handbook of emotion elicitation and assessment},
  pages={106--123},
  year={2007},
  publisher={Oxford University Press New York, NY}
}

@article{robbins2014,
  title={Cancer conversations in context: naturalistic observation of couples coping with breast cancer.},
  author={Robbins, Megan L and L{\'o}pez, Ana Mar{\'\i}a and Weihs, Karen L and Mehl, Matthias R},
  journal={Journal of Family Psychology},
  volume={28},
  number={3},
  pages={380},
  year={2014},
  publisher={American Psychological Association}
}

@article{russell1980,
  title={A circumplex model of affect.},
  author={Russell, James A},
  journal={Journal of personality and social psychology},
  volume={39},
  number={6},
  pages={1161},
  year={1980},
  publisher={American Psychological Association}
}

@article{schoebi2008,
  title={The coregulation of daily affect in marital relationships.},
  author={Schoebi, Dominik},
  journal={Journal of Family Psychology},
  volume={22},
  number={4},
  pages={595},
  year={2008},
  publisher={American Psychological Association}
}

@article{settineri2014,
  title={Caregiver's burden and quality of life: Caring for physical and mental illness},
  author={Settineri, Salvatore and Rizzo, Amelia and Liotta, Marco and Mento, Carmela},
  journal={International Journal of Psychological Research},
  volume={7},
  number={1},
  pages={30--39},
  year={2014},
  publisher={Facultad de Psicolog{\'\i}a. Universidad de San Buenaventura, Medell{\'\i}n}
}

@article{steyer1997,
  title={Der Mehrdimensionale Befindlichkeitsfragebogen MDBF [Multidimensional mood questionnaire]},
  author={Steyer, Rolf and Schwenkmezger, Peter and Notz, Peter and Eid, Michael},
  journal={G{\"o}ttingen, Germany: Hogrefe},
  year={1997}
}

@article{watson1988,
  title={Development and validation of brief measures of positive and negative affect: the PANAS scales.},
  author={Watson, David and Clark, Lee Anna and Tellegen, Auke},
  journal={Journal of personality and social psychology},
  volume={54},
  number={6},
  pages={1063},
  year={1988},
  publisher={American Psychological Association}
}

@article{zepf2020,
  title={Driver emotion recognition for intelligent vehicles: a survey},
  author={Zepf, Sebastian and Hernandez, Javier and Schmitt, Alexander and Minker, Wolfgang and Picard, Rosalind W},
  journal={ACM Computing Surveys (CSUR)},
  volume={53},
  number={3},
  pages={1--30},
  year={2020},
  publisher={ACM New York, NY, USA}
}
